# Gamma-ray imaging system for real-time measurements in nuclear waste characterisation

L. Caballero [1,*], F. Albiol Colomer [1], A. Corbi Bellot [1], P. Olleros Rodríguez [1], J. Agramunt Ros [1], C. Domingo-Pardo [1], J.L. Leganés Nieto [1], P. Contreras [1] and D.L. Pérez Magán [1]

[1] *Instituto de Física Corpuscular - CSIC - University of Valencia; C/Catedrático José Beltrán, 2; E-46980; Paterna; Spain*
[1] *ENRESA, Spain*
  *E-mail*: Luis.Caballero@ific.uv.es

ABSTRACT: A compact, portable and large field-of-view gamma camera to identify, locate and quantify gamma-ray emitting radioisotopes in real-time has been developed. The device delivers spectroscopic and imaging capabilities, which allow one to use it in a variety of nuclear waste characterisation scenarios, such as radioactivity monitoring in nuclear power plants and more specifically for the decommissioning of nuclear facilities. The technical development of this apparatus and some examples of its application in field measurements are reported in this article. The performance of the presented gamma-camera is also benchmarked against other conventional techniques.

- KEYWORDS: Radiation monitoring; Inspection with gamma rays

---

* Corresponding author.

# Contents

## 1. Introduction and motivation

Gamma spectrometry is extensively used in radioactive characterization of nuclear waste and in the survey of contaminated areas in Nuclear Power Plants (NPP). However, this type of systems, either portable or stationary, typically have a low detection efficiency and require rather long measuring times until the final diagnostic becomes available. This is particularly evident where large areas (walls, containers, etc.) have to be manually *scanned.* These scanning procedures are usually performed by means of traditional counters and/or collimated gamma-ray detectors. Hence, a detecting system with a large field of view (FOV), becomes of special interest.

Another motivation for the development and use of such an imaging system arises from the fact that many times large volumes of radioactive waste are sorted according to their volume- or mass-averaged activity. However, there are situations where most of the activity is concentrated in one single isolated region of that large volume. By quickly identifying such scenarios, it would become possible to optimize the classification procedures, and eventually reduce the total amount and costs for nuclear waste disposal. In general, the more precise the spatial distribution of the activated regions in walls, containers and drums is known, the better the decontaminating or dismantling strategy could be defined. In summary, many important parameters as dismantling resources, budget, resources and operator radioactive exposure could be minimized with a faster and more complete and detailed knowledge of the radioactive scenario.

Usual procedures in radioactive contamination characterization are usually manually performed with radiation meters by the NPP operators. There are situations where such approach might be very time consuming, not sufficiently accurate and implies large operator exposure to radiation. Several gamma-imaging systems have been developed and are now commercially available [1-4]. These devices allow to locate radioactive sources and determine the contamination of the scenario in shorter times, thus reducing the operator exposure dose received. However, there is still room for improvements regarding portability and accuracy of the imaging system to complement and meliorate the current procedures.

In that sense, we have developed GUALI (Gamma Unit Advanced Location Imager) in collaboration with ENRESA, the Spanish National Agency for the Management of Nuclear Waste. GUALI represents a versatile, compact and portable gamma-ray imaging system. It allows operators to map radioactive sources in contaminated environments, as well as precisely determine the radioactive contamination distribution and activity. A fundamental distinctive feature of GUALI when compared with other systems commercially available resides on its capability to geometrically recognise the environment by means of a coupled optical system. This system is in turn based on a conventional RGB camera and modern computer vision techniques [5]. During each measurement GUALI continuously displays a superposition of the image for the measured gamma-activity spatial distribution together with the optical one, thus aiding the quick identification and location of the radioactive sources in the measurement scenario. Furthermore, GUALI is capable of automatically identifying a movement or a change in the image-plane, thus triggering its own (image and gamma) acquisition systems accordingly, saving data and RGB images consistently, hereby minimizing human mistakes during the decommissioning works.

This article is structured as follows. Section 2 provides an overview of the GUALI system and a succinct technical description of its main elements. In Section 3 both the spatial- and energy-calibrations are described, together with the methodology developed to account for certain experimental effects. Sections 4 and 5 report several examples to illustrate the performance of the present apparatus, in controlled laboratory measurements and in field measurements at a NPP, respectively. Conclusions are drawn in Section 6.

## 2. System design

GUALI consists of a pinhole gamma camera for detecting gamma-rays supplemented with an optical RGB-camera. The latter will account for movement detection and system tracking, as will be explained in Section 2.3. Both imaging systems are mounted on a portable cart, which also accommodates the control electronics, power supply and a compact computer to process and analyse the acquired data, and to display the results on the screen. An integrated Wi-Fi module allows the operator to connect remotely to the system and display the results on a web browser using a portable device such as a Tablet or Laptop. This aspect allows the operators to work at larger distances, thus being particularly useful to reduce their exposure and dose-budget. The GUALI system design allows operators to easily move the system and perform measurements at different areas of the facility.

### 2.1 Gamma detection head

The system detection head is comprised of a position-sensitive gamma-ray detector coupled to a pinhole collimator manufactured in lead, restricting radiation impinging in the gamma detector to have passed through the pinhole, thus enabling image reconstruction. The collimator thickness has been calculated by means of the Geant4 Monte Carlo (MC) simulation code [6] to shield from the gamma-rays expected in the NPP scenario (see details in Section 2.2).

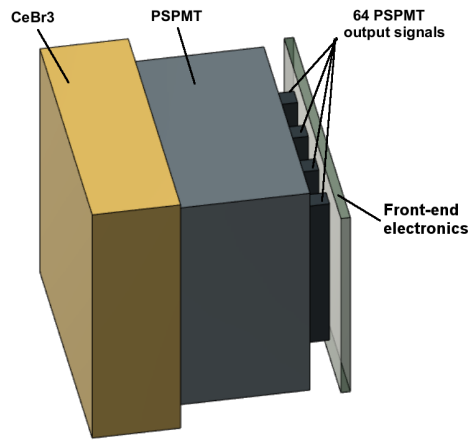

**Fig. 1: Schematic view of GUALI detection head comprising the gamma detector parts.**

The gamma-ray detector consists of a continuous $CeBr_3$ scintillator crystal of $51\times51\times15mm^3$ optically coupled to a position-sensitive photomultiplier tube (PSPMT) from Hamamatsu Photonics (H10966A100). This PSPMT features a photocathode area of $49\times49mm^2$ and a nominal quantum efficiency of 35% at peak wavelength of 400 nm. The photosensor has 64 anode segments with a nominal photo-electron gain of $35\times10^5$, and anode gain uniformity factors of up to 50%. The fact of reading out and processing individually all 64 signals would become expensive in terms of cost and system portability. There are several options commercially available based on resistor networks [7]. However, for availability reasons and because they are expected to perform better in terms of uniformity corrections, we used the application specific integrated circuit (ASIC) AMIC2GR [8, 9, 10], which has been mainly developed for medical PET applications. The AMIC2GR reads 64 inputs from the PSPMT (see Fig. 1) and outputs up to 8 analogue signals (called moments), which correspond to the arithmetic sum of all 64 pixels, each of them weighted by a pre-programmable weight. The methodology employed to reconstruct positions from the moments is similar to the working principle of an Anger camera [11]: five different coefficient matrices are required to recover the energy and position of detection (X, Y) of each gamma-ray event on the PSPMT surface. The energy E is obtained from the first moment M0, for which is applied a uniform matrix (all coefficients are equal) corrected by an additional matrix M, that counterbalances the inhomogeneity in response between the pixels of the photocathode (as provided by the manufacturer). In GUALI, the gamma-ray interaction position (X,Y) in the crystal plane is recovered from four different momenta, A, B, C and D, corresponding to diagonal weighting-gradients, also corrected by the inverse of the pixel-gain uniformity factors. The methodology is similar to the one described in [12].

## 2.2 Pinhole collimator

Planar gamma cameras are based on the use of collimators to determine the spatial origin of the detected radiation. Among the different collimator types and designs, a convenient balance between spatial resolution and detection efficiency must be achieved. In our case, given the average activities (1-100 Bq/g) of the objects for inspection, a pinhole collimator type has been found convenient for a number of scenarios in NPPs and for typical decommissioning activities. The aim of the pinhole collimator is to shield the scintillator crystal from any gamma radiation not directly coming through its aperture. Several simulations using the Geant4 MC toolkit have been carried out to determine the shielding material, thickness, pinhole geometry and dimensions, with the aim of achieving a reasonable compromise between detection efficiency and spatial resolution. Lead was chosen as shielding material and its thickness was calculated to attenuate at least 85% from 1332 keV gamma-rays coming from any direction in the forward hemisphere. The dimensions of the detector FOV are determined by construction and depend on the pinhole focal length. In our case, the collimator was designed with a diameter of 3 mm and an aperture angle of 24 degrees. The distance between the centre of the pinhole and the front face of the scintillation crystal is 40 mm. Since we do not have the information of the depth of interaction of the gamma-ray inside the crystal, from the MC simulation results we can assume that gamma-rays are detected in a plane situated in the middle of the crystal (7.5 mm from the crystal frontal face). MC simulations showed an average depth of interaction of 7.5 mm (9.5 mm) for 662 keV (1332keV). The reconstructed images for both 47.5 and 49.5 mm did not show significant differences. Hence, our focal length was assumed to be 47.5 mm for all the image reconstruction.

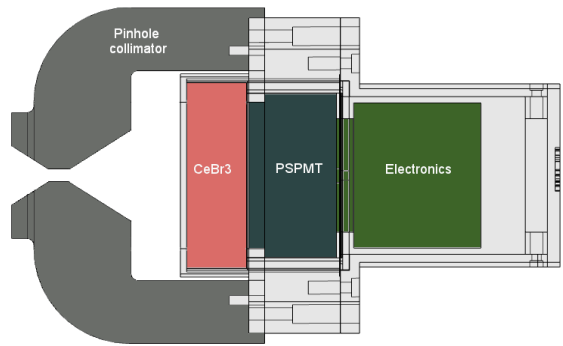

**Fig. 2: Detection head parts outlines.**

## 2.3 Optical images

An optical RGB- camera (1280x720 pixels) has been mounted over the gamma detection head (see Fig.3). Its orientation is roughly parallel to the gamma camera in order to be able to warp, overlap and finally alpha-blend (overlapping and composition with a degree of transparency) a colour snapshot (plain photograph) over its *gammagraphic counterpart* (or vice-versa). In order to do so, the dual-camera system must undergo a calibration phase, which is detailed in [13]. During this phase, a set of gamma measurements from point-like radioactive sources and their corresponding RGB images were obtained. Their goal is to accurately find the rigid transformation (rotation+translation) that connects both optical systems (i.e., both pinholes).

Once the system is geometrically calibrated, the aforementioned multi-modality image overlapping can take place. With more detail, white-and-black markers composed of binary patterns (visible fiducials) are randomly attached by the operator around the object under inspection (see below Fig. 16). These markers are automatically and continuously identified and tracked by means of the methodology/algorithm described in [14, 15]. The outcome of this steady marker identification is a 3D model of the inspected object surface in the FOV of the visible camera. The geometry of this 3D model can then be rigidly translated to the coordinate space of the gamma camera. In normal operation mode, the system is then suited for calculating the distance to any visible marker and can provide a good overlap between optical and gamma images.

## 2.4 Acquisition electronics and cart

Signals from the detection head electronics are collected and processed using a fast-sampling digitizer (SIS3316-DT from Struck Innovative Systeme GmbH), which features a sampling rate of 250 MS samples/s and a 14-bit range ADC. A computer records and processes the acquired data with a customized software developed at the Gamma-Ray Spectroscopy and Neutron group of IFIC [16]. All these components, together with the DC-power supply, are allocated inside a portable cart designed to be safely and easily moved through the different measurement and diagnostic stations of the NNP facility.

In order to aid the operators to correctly position the system, a 50 mW laser pointer is also embedded in the detector head. A small rotating arm allows to point the laser beam in collinear coincidence with the main axis of the FoV of the gamma-camera. Once the system is positioned and before starting the acquisition, the laser is removed to the lateral position (Fig.3).

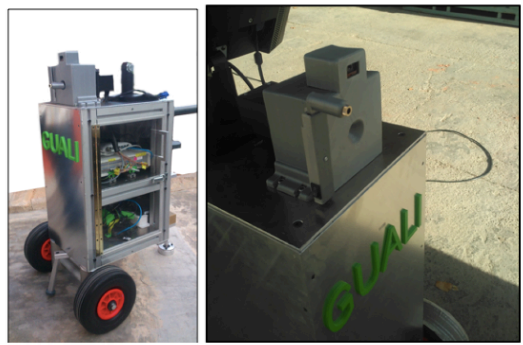

**Fig. 3: GUALI imaging system cart containing the acquisition electronics (left) and detection head close up with the movable laser to indicate the centre of the FOV (right).**

## 3. System performance

In the following Subsections, we describe the calibrations and the measurements carried out in order to characterize the performance of the system. Two of the most abundant gamma-ray emitting radionuclides in the NPP scenarios are $^{137}$Cs ($t_{1/2}$=30.1y) and $^{60}$Co ($t_{1/2}$=5.27y). Thus, in order to calibrate the system in the energy range of interest we have measured the energy pulse-height spectrum of three different point-like calibration sources ($^{22}$Na, $^{137}$Cs, $^{60}$Co) placed at 5 cm from the centre of the pinhole. The energy calibration was accomplished by adjusting a Gaussian distribution to the full energy deposition events. The detector shows a linear behaviour in the whole energy range, and yields an energy resolution of 7.6% FWHM at 662 keV. For the diagnostic purposes of GUALI, this energy resolution is enough since it allows one to distinguish the two most common radioisotopes expected in NPP scenarios ($^{137}$Cs and $^{60}$Co).

### 3.1 Position reconstruction

Scintillation photons produced by an impinging gamma-ray in the scintillation-crystal volume spread isotropically and are collected in the photocathode surface of the PSPMT. The maximum of this light distribution in the photocathode corresponds to the X, Y coordinates of the event position. However, the light from events close to the crystal border suffers higher reflection in the lateral surfaces of the crystal and thus, the centre of the light distribution is displaced toward inner coordinates. In order to correct for this pin-cushion or border effect in the spatial reconstruction and recover the original centroid, a position calibration must be performed.

Two different methods were explored in the present work for the position reconstruction calibration. In the first method, a collimated gamma source was moved along the front detector face at known coordinates and their raw images reconstructed. The source is displaced along the plane of the detector surface by intervals of 6 mm horizontally and vertically forming an 8×8 spatial grid, centred on the detector. The second method consisted in reconstructing the images of a gamma-ray source at different positions along the detector FoV in a focal plane at a known distance. In this case, the source was moved by intervals of 4 cm in vertical and horizontal directions, thus forming a 9×9 spatial grid. The focal plane used was at a distance of 50 cm from the pinhole.

In both cases, after reconstructing the acquired images in all known positions, a 2-dimensional fit is performed to find the two-dimensional polynomial parameterization between the raw measured ($x',y'$) positions and the real or true positions $(x,y) = (f_1,f_2)$. To this aim, the class TMultiDimFit implemented in the ROOT [17] software package was used. This class is based on the combination of the CERNLIB MUDIFI package [18] and MINUIT [19], which improves the fitting errors by means of the class TMinuit. Following the methodology described in [20] we define two functions $f_1 = f_1(x',y')$ and $f_2 = f_2(x',y')$ given by

$$f_1(x',y') = \sum_{i=0}^{M} E^i (x')^{a_i} (y')^{b_i}$$

$$f_2(x',y') = \sum_{i=0}^{N} F^i (x')^{c_i} (y')^{d_i}$$

and find the values of all the parameters $\{a_i, b_i, c_i, d_i, E_i, F_i\}$ such that $(x_R \equiv f_1(x',y'), y_R \equiv f_2(x',y'))$ correspond to the best fit of the data sets of points $\{(x_i, y_i), i = 1 \ldots 64\}$. The number of terms M and N for the reconstructed positions f1, and f2 was of 12. This result was found after progressively increasing these quantities, until the result was found to not significantly improve with the addition of further terms. Average (maximum) deviations of 0.14 (0.3) and 0.3 (1.3) mm with respect to the *x* and *y* real positions were found.

Measured and reconstructed positions after applying the function for both methods are shown in (Fig.4). Reconstructed position errors were very similar for both methods. We decided to use the second method since it contained more experimental data points and it already accounts for possible distortions introduced by the pinhole-collimator itself. The system showed a very good linearity in the central $30\times30$ mm$^2$ area of the detector focal plane. This area corresponds to a Useful Field of View (UFOV) of approximately $2\times2$ m$^2$ at a focal plane placed 3 m in front of the detector pinhole. Such distance can be considered representative of the spatial distance commonly used to explore nuclear waste containers.

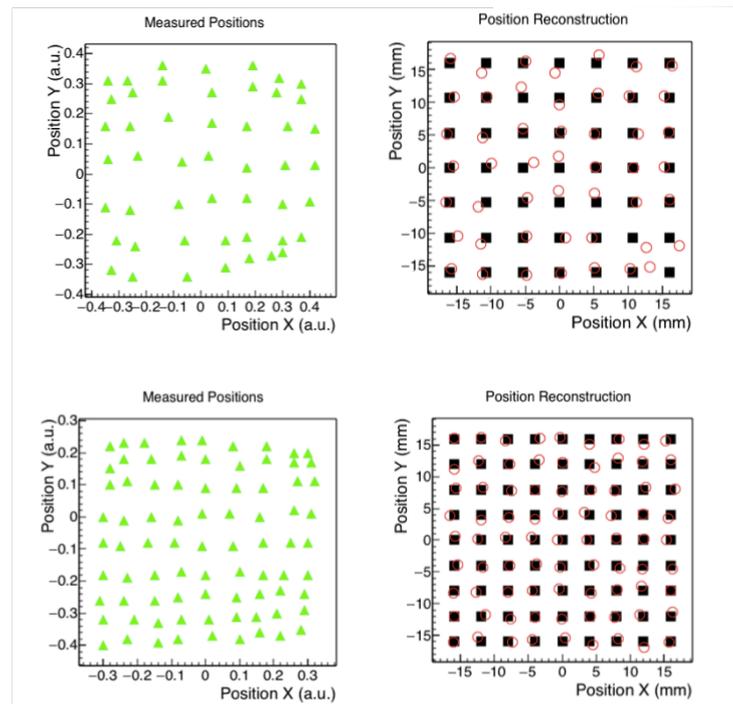

**Fig. 4: Green solid triangles represent the measured positions (left panels). The reconstructed positions are displayed on the right panels by open circles. Results for the collimated gamma source along the detector front face method are shown in the top panels, whereas the method using the gamma source along the detector FoV at a known focal plane is illustrated in the panels below. Solid square symbols represent the true positions in each case.**

### 3.2 Uniformity correction

To identify and properly quantify the activity of radioactive sources represents a fundamental aspect for both nuclear waste classification and inspection of contaminated areas. For this reason, besides the image compression correction mentioned in the previous section, image uniformity is the other basic parameter. An imaging procedure of a homogeneous flat source must produce a homogeneous gamma image, i.e., no artificial activity "hot spots" should appear in the image. Despite of the afore described correction applied in the response of the 64 pixels, there are additional distortions introduced by the imperfections of the pinhole collimator itself. In order to understand our detector response and to correct for this effect, we have applied two different methods. In the first method, the detector geometrical response has been calculated through a

– 7 –

Monte Carlo simulation using the Geant4 code based program. The detector was modelled in the code using a realistic description of it, which includes its exact geometry for the scintillator crystal, collimator, etc. The event generator comprises a flat homogeneous $^{22}$Na source covering the full FOV of the detector. A $^{22}$Na source was found convenient for this study because its two gamma-rays (511 and 1275 keV) are covering, in a single source, the main energy range of the expected radioisotopes in typical NPP scenarios, such as $^{137}$Cs and $^{60}$Co. In the MC simulations, the electromagnetic interactions were modelled using the G4EMLOW6.48 Low Energy Geant4 package. For each gamma-ray interacting within the crystal volume, both energy and position of the gamma-ray interactions were recorded for posterior analysis. Scintillation photons and optical light reflection and refraction effects were not simulated, as we are interested in obtaining the reference spatial response that will be used to correct the measured one.

On a second method, measurements have been performed in order to experimentally find the geometrical uniformity correction. A $^{22}$Na source of 1 mm size and 530 kBq activity was placed in different positions along the focal plane at 50 cm of distance from the pinhole. The source was moved in discrete steps along the X and Y axis covering all the UFOV. For each measurement, its image was reconstructed as described in Section 3.2. Acquisition duration in all the positions was the same in order to statistically compare the photopeaks statistics of the reconstructed images. These results were used to calculate the experimental geometrical uniformity correction. A smoothing (using kernel algorithms) on the 2D discrete measured values for comparison with the simulated ones has been applied for both the 511 and 1275 keV $^{22}$Na photopeaks uniformities. More specifically, in our case we have iteratively applied 5 times the smoothing filter based on a kernel acting on 5×5 cells. In Fig.5. plots of the simulated and measured uniformities are shown.

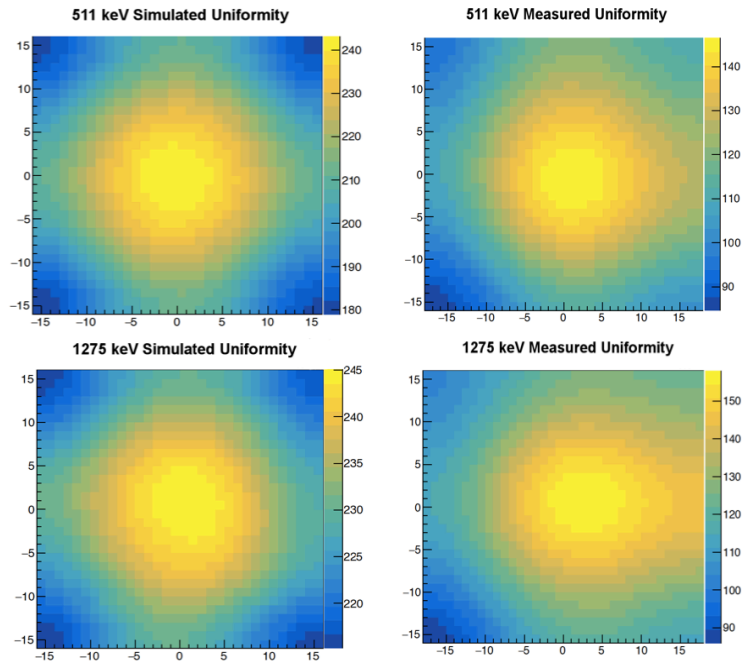

**Fig. 5: Simulated (left) and measured (right) uniformities (in arbitrary units) for 511 keV (top) and 1275 keV (bottom) gamma-rays.**

The results obtained (see Fig.5) reproduce, in general terms, the expected behaviour of the inverse square law for the radiation passing through the pin-hole collimator, thus showing a radial gradient for the intensity registered at each position of the sensitive surface. However, while a symmetry around the centre is predicted by the simulation, in the measured distributions one can clearly appreciate an anisotropy in the X axis for both energies. This asymmetric response could be ascribed so several experimental effects, such as to an inhomogeneous response of the scintillation crystal itself, inhomogeneous optical coupling or a small misalignment (estimated about 1.5 mm) of the pinhole with respect to the centre of the crystal. Regardless of its origin, the measured response at 511 and 1275 keV is used to correct for spatial inhomogeneity in $^{137}$Cs and $^{60}$Co, respectively. In order to apply the uniformity correction, we used a numerical approach based on a lookup table with a pixel-by-pixel correction for new measured values (shown in the right column of Fig.6).

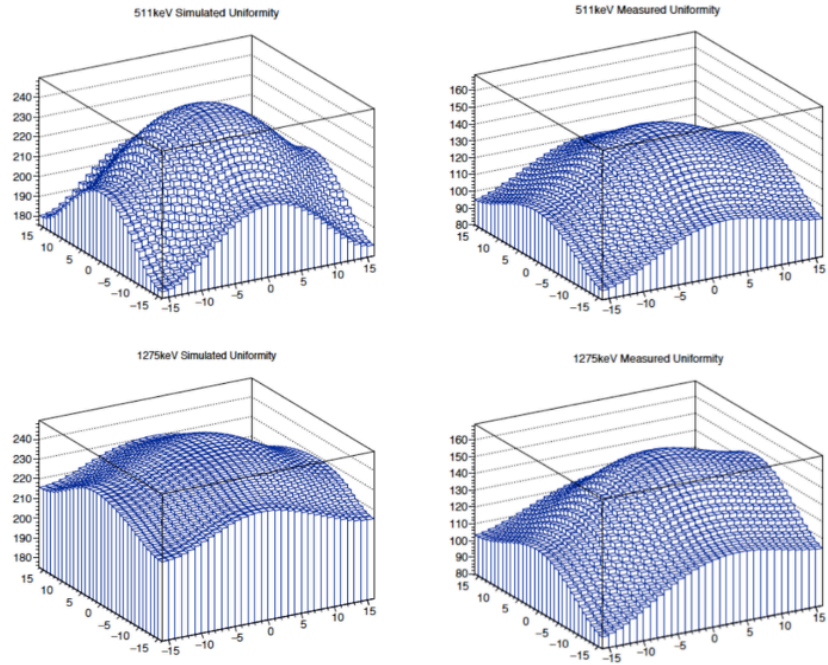

**Fig. 6: Simulated (left column), measured (right column) of the measured uniformities for 511 keV (above) and 1275 keV (below) gamma-rays.**

In order to evaluate the quality of the uniformity correction a set of measurements with identical duration of a $^{137}$Cs source positioned in nine positions in the same focal plane were carried out. Thus, this set of measurements simulates a 3×3 array of sources of equal activity. The look-up table correction described above was applied to these measurements. As it can be seen in Fig.7, the intensities of the sources are reduced to deviations lower than a 10% over the UFOV relative to their average value once the uniformity correction is applied.

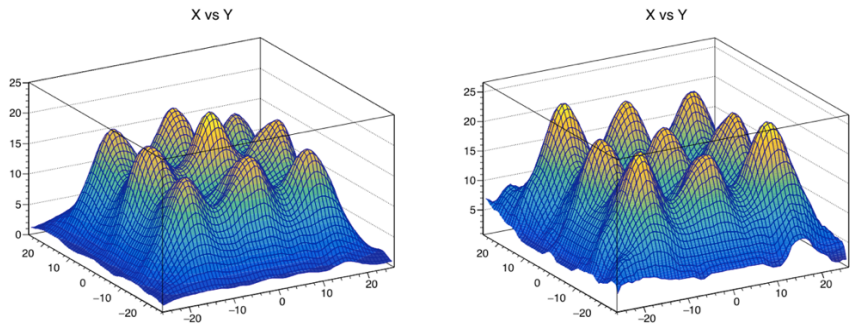

**Fig. 7: Reconstructed images of an array of 9 sources before (left) and after (right) applying the uniformity correction.**

### 3.3 Acquisition electronics and cart

Since many different radioisotopes might be present in the scenarios where GUALI is used, the energy spectrum is composed of both the Compton- and the Full-energy-event contributions from all the radioisotopes present in the detector FOV. For each radionuclide, the gamma-ray image is obtained by putting an energy range condition in the full energy spectrum so that only events registered inside that energy range or gate are used to produce the corresponding gamma image. The gamma activity for each radioisotope is obtained by integrating only the events included in the corresponding photopeak. However, the Compton spectra generated by higher energy photopeaks from other radioisotopes contributions inside of the selected gate will degenerate the intended photopeak image. In order to subtract this background contribution, a similar gate in width is placed just on the right side of the photopeak, from which we want to produce the image. After reconstructing both images, from the photopeak and the Compton suppression gates, we subtract the latter from the former, so that the resulting *clean* image contains mostly the desired radioisotope photopeak contribution. This approach is illustrated in Fig.8, where one can observe the different gates placed in order to form the $^{137}$Cs and $^{60}$Co gamma-ray images. In the $^{137}$Cs case, the 662 keV photopeak gate has a contribution coming from the $^{60}$Co Compton continuum. In the $^{60}$Co case, the background suppression gate is as large as the sum of the two gates placed at 1173 and 1332 keV photopeaks.

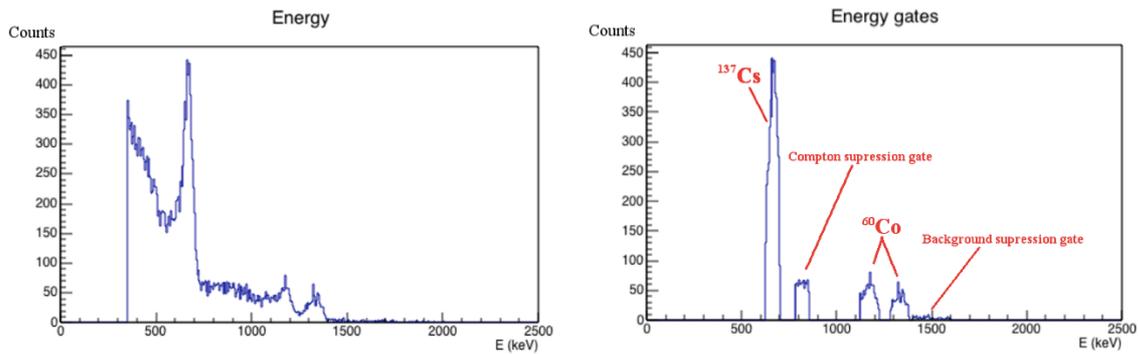

**Fig. 8: Energy spectrum (left) and gates placed to form the reconstructed images for $^{137}$Cs and $^{60}$Co (right).**

### 3.4 Activity determination and system sensitivity

Besides knowing the spatial distribution of the gamma radioisotopes in objects or surfaces, it is also important to determine and quantify the activity of the identified gamma-ray sources in the field of view. However, since the images obtained by the gamma detector are formed using a pinhole collimator (2D-sensitivity), it is not possible to directly determine the depth of the gamma-ray emitting source. Alternatively, it is possible to determine the apparent or the effective source activity at a given surface, located at a certain distance from the detector. This quantity corresponds to the value that an operator could measure with a counter or detector at contact with the corresponding volume surface. To this aim, GUALI makes use of its RGB camera and external binary visual fiducials, which were introduced before in Section 2.3. The software computes the distance between the visual markers and the pinhole of the gamma camera, which is then used as the system's focal length. The accuracy of determining the distance with the Aruco frames depends on the number of frames used, being at least of 1 cm at 1 m distance when only one marker is used.

In order to quantify the registered activity, the gamma camera must be calibrated in efficiency. In order to characterize the efficiency, different radioactive sources with known activities were placed in the centre of the FOV at a known distance (19 cm). In our case, $^{22}$Na, $^{137}$Cs and $^{60}$Co sources were used for this calibration. Results for the full-energy efficiency are shown in Fig.9 together with the values obtained from the simulations with the Geant4 simulation code. An efficiency of 0.0046±0.0002% was measured for the 662 keV photopeak. As expected, the trend of the values is similar and there is a discrepancy of the 30% between the simulation and the measured values. This disagreement can be mainly ascribed to imperfections in the geometrical modelling of our detection apparatus, both the crystal itself and the massive lead collimator.

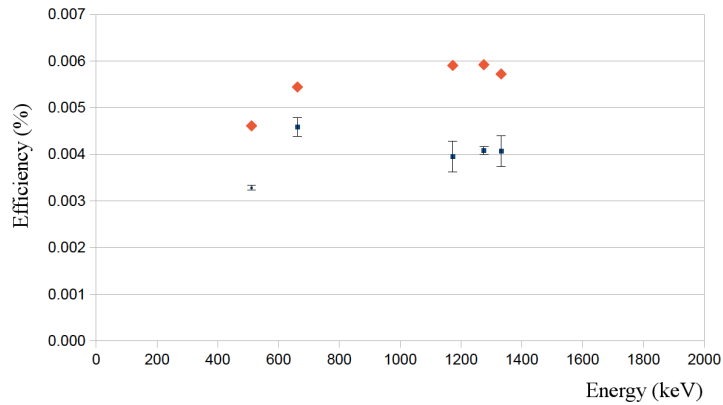

**Fig. 9: Simulated (red) and measured (blue) detection efficiency as a function of the energy for sources placed at 19 cm from the detector.**

### 3.5 Spatial resolution

A determination of the spatial resolution of the system becomes necessary in order to have an estimation of the in-field gamma detector resolving power of the radioactive sources being imaged. The experimental setup consisted of systematic measurements of two $^{137}$Cs sources with a variable separation ranging from 10 to 16 cm in 2 cm steps, in the focal plane at 50 cm from the pinhole. Reconstructed images for each of the positions are shown in Fig.10. As can be observed, the sources are resolved in the images when they are separated at least by 12 cm. Thus, in the NPP scenario, where the object under inspection is approximately at 3 m, GUALI will be capable of resolving radioactive sources of similar activity separated by at least 72 cm.

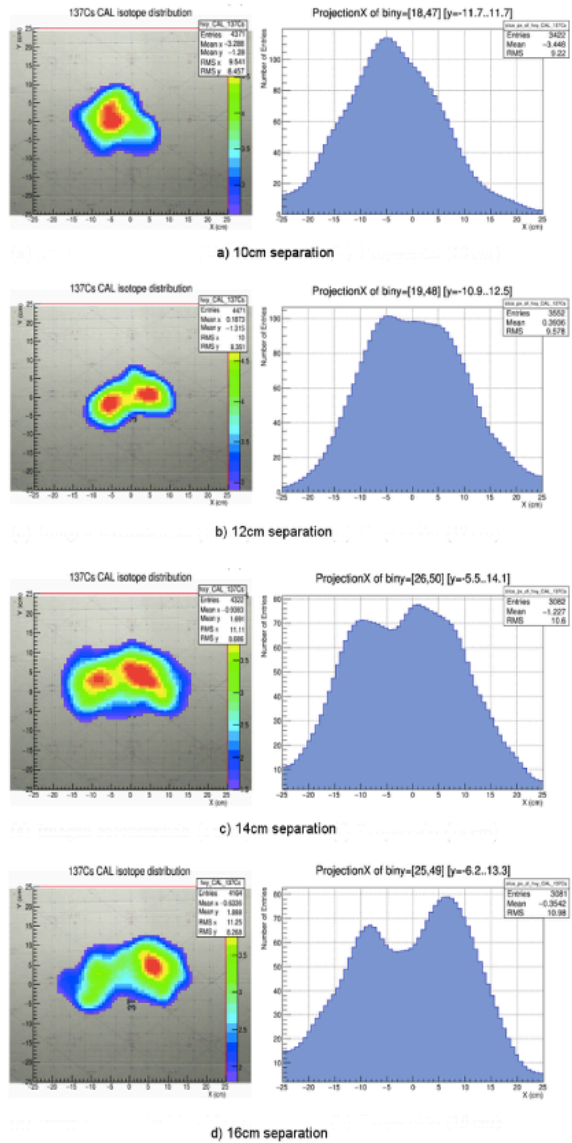

**Fig. 10:** Reconstructed images and X-projection for sources separation of 10, 12, 14 and 16 cm at 50 cm from the pinhole.

## 4. Lab measurements

After calibrating and characterizing the system, a set of measurements were carried out in order to evaluate the performance of GUALI. Laboratory measurements were designed and performed using radioactive sources, in order to mimic localization and identification scenarios similar to those encountered in NNPs. The objective was both to check whether the system was suited for a correct image reconstruction and if the correspondence of gamma and optical images was properly done. The first test performed was to place a 0.53 MBq $^{22}$Na radioactive source in different positions in a wall and the GUALI system 70 cm away. As can be observed in Fig.11, the gamma image of the radioactive source clearly matches the position in the optical one.

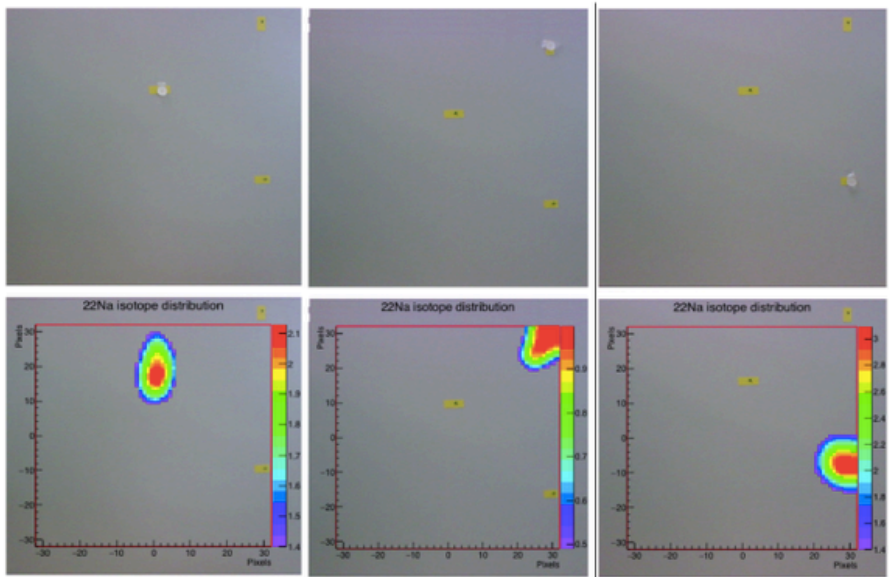

**Fig. 11:** **$^{22}$Na radioactive source imaging in different positions in a wall at 70 cm of distance from the detector. Above, optical images. Below, gamma superimposed to the optical images.**

A similar set of measurements (shown in Fig. 12) was performed increasing the detector to focal plane distance up to 250 cm, which is a more realistic distance for imaging the nuclear waste containers. Again, a nearly perfect correspondence of the reconstructed image for the gamma sources and their optical images is obtained.

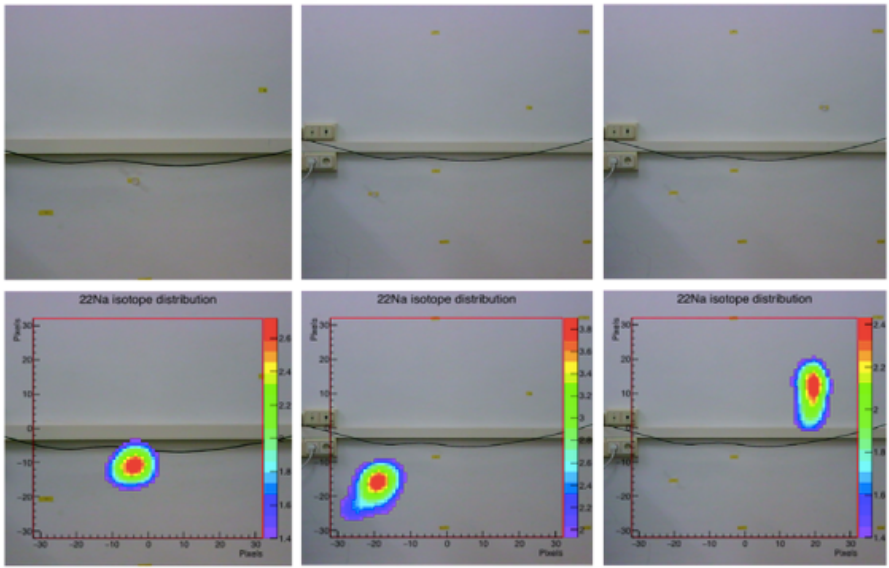

**Fig. 12:** **$^{22}$Na radioactive source imaging in different positions in a wall at 250 cm of distance from the detector. Above, optical images. Below, gamma superimposed to the optical images.**

## 5. In-field measurements

After characterizing the system in the laboratory, GUALI was moved to a real operation scenario to image decommissioning objects with usual activities and radio-contaminants contents in order to evaluate its imaging capabilities and compare them directly with the standard procedures. The measurements were carried out at the José Cabrera Nuclear Power Plant sited in Zorita, Guadalajara (Spain), which is currently under decommissioning. At this NPP, nuclear waste from the dismantling works is sorted according to their activities in containers and drums, which are labelled and stored depending on their average activity per kg. Currently, nuclear waste classification is based on a sequential spectroscopic scan of the waste containers with collimated germanium detectors, used as a box-counter to register the total activity in their FOV. The system used is a WM 2500 Modular Gamma Box and Container Assay System commercialised by Canberra consisting in four High-Purity-Germanium detectors mounted on two opposite columns aside a mobile platform in which the container is placed [21], capable of detecting activities over 0.01 Bq/g in CMB containers. The collimation aperture for the HPGe detectors is adapted to the total volume object of the scan. As can be seen in Fig.13, four germanium detectors are placed in two opposite towers, while the container is positioned in between them. Typically, two - three positions are measured (corresponding to eight - twelve single Ge-spectra) in order to fully scan one container. Each measurement typically lasts for 5 min for containers, thus being the total characterization time of 10 – 15 minutes per container.

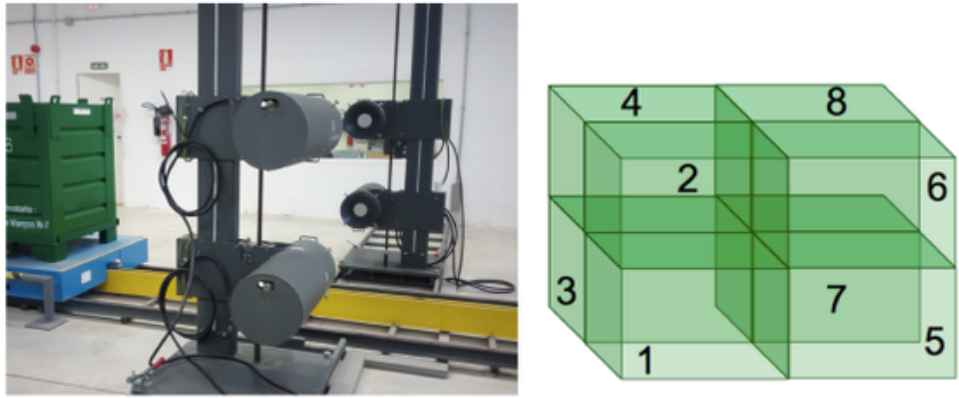

**Fig. 13: Box-counter for container activity measurement (left) and sub-volumes scheme (right).**

| Radionuclide | Derived segment activity from the sum spectrum $S_{SUM}$ (Bq/g) | | | | | | | | $S_{SUM}$ (Bq/g) |
|---|---|---|---|---|---|---|---|---|---|
| | S1 | S2 | S3 | S4 | S5 | S6 | S7 | S8 | |
| $^{137}$Cs | 16.6 | 20.1 | 11.2 | 12.4 | 47.5 | 59.0 | 9.4 | 17 | 24.1 |
| $^{60}$Co | 7.9 | 5.5 | 8.0 | 5.9 | 33.8 | 12.3 | 7.6 | 3.7 | 10.6 |

**Table 1: Derived segment activity from the sum spectra (Bq/g) measured in the box-counter.**

With this technique, the box-counter registers the activity of the container in eight subvolumes as shown in Table 1 and in the schematic activity distribution view in Fig.16. The total activity registered in this CMB-type container and its weight gives the container activity in Bq/g. For this container, the total average activity was of 24.1 Bq/g and 10.6 Bq/g for $^{137}$Cs and $^{60}$Co, respectively. This parameter is used to determine whether the container can be classified as Very Low Level Container or released from regulatory control. Activity in each sub-volume is derived from the total sum spectrum accumulated from all the detectors. However, by means of this approach it is only possible to identify hot-spots between sub-volumes. Each sub-volume contribution is smeared out in the FOV of each HPGe detector measurement.

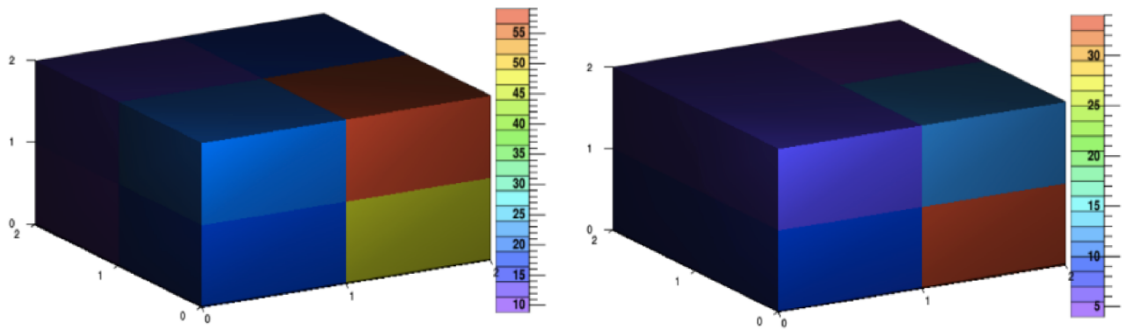

**Fig. 14: Container activity measured with the box-counter for $^{137}$Cs (left) and $^{60}$Co (right) radionuclides. Total activity (in Bq/g units) registered is spread over eight subvolumes.**

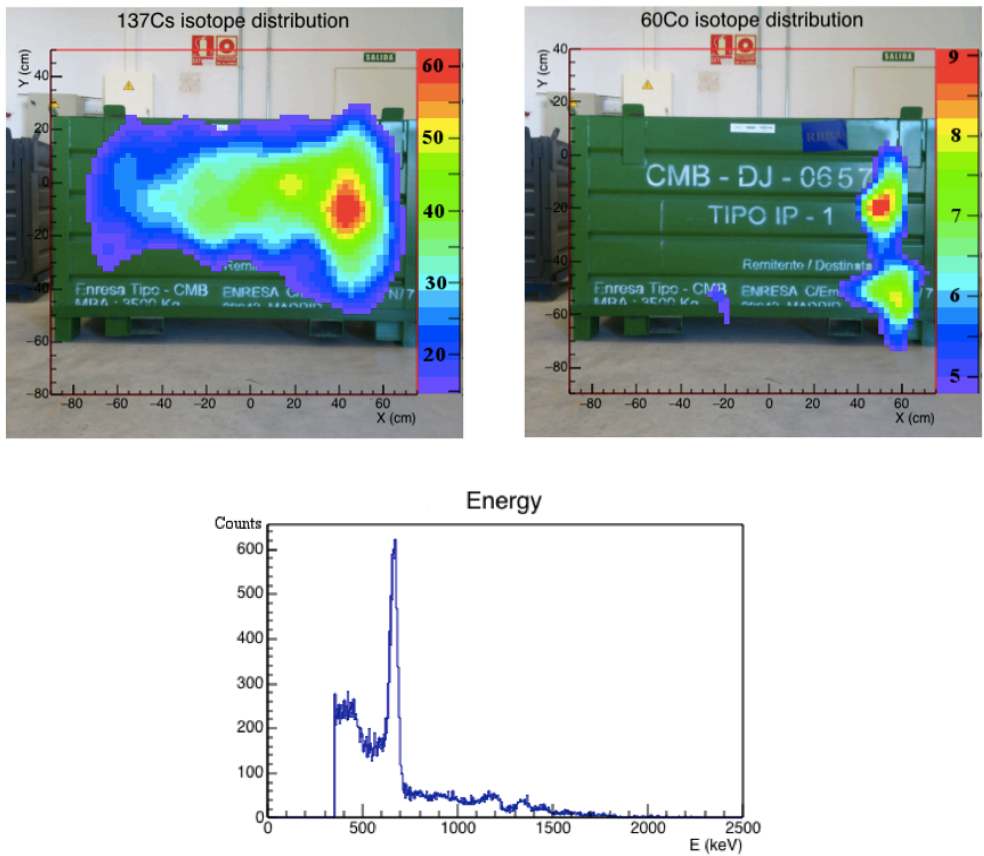

**Fig. 15: Gamma images of the container activity distribution obtained with GUALI for $^{137}$Cs (top left) and $^{60}$Co (top right) radioisotopes (in Bq/g units). Energy spectrum is shown in the bottom.**

Fig.15 shows in-field measurement carried out with GUALI for the same container analysed with the box-counter (Fig.13-14 and Table 1). The duration of acquisitions was 30 minutes. Considering that GUALI measure is only one side of the container, the total acquisition time per container is 4 times larger than a 4 HPGe-based box-counter. During the acquisition and according to the energy spectrum being acquired, the operator can define energy windows on the photopeaks of interest and obtain their corresponding gamma images. For this container, the total average activity measured was of 21.9±0.5 and 7.3±0.4 Bq/g for $^{137}$Cs and $^{60}$Co, respectively. Both activity values were in good agreement (quoted uncertainties are statistical) with the previously measured in the box-counter. When comparing the measured spectra, the box-counter shows a much higher energy resolution, sensitivity and spectroscopic capabilities than GUALI due to the use of HPGe, whereas in terms of spatial distribution, GUALI offers a much more detailed information. Both systems form a very compatible and complementary system for both spectroscopic and imaging characterization of nuclear waste containers. With all this information available, operators can decide whether a material reordering is necessary or not.

The use of the visual markers to match the optical and gamma images is shown in Fig.16, in which both $^{137}$Cs and $^{60}$Co distributions are shown. As previously mentioned, during acquisition GUALI is continuously registering the visual information and calculating the average plane in which they are attached, which is used as the object focal plane in the gamma image. Additionally, the software is capable of detecting small changes in the optical image, and automatically stop the current gamma acquisition session. The acquisition is then left in stand-by until a new static situation is measured, thus initiating a new data acquisition.

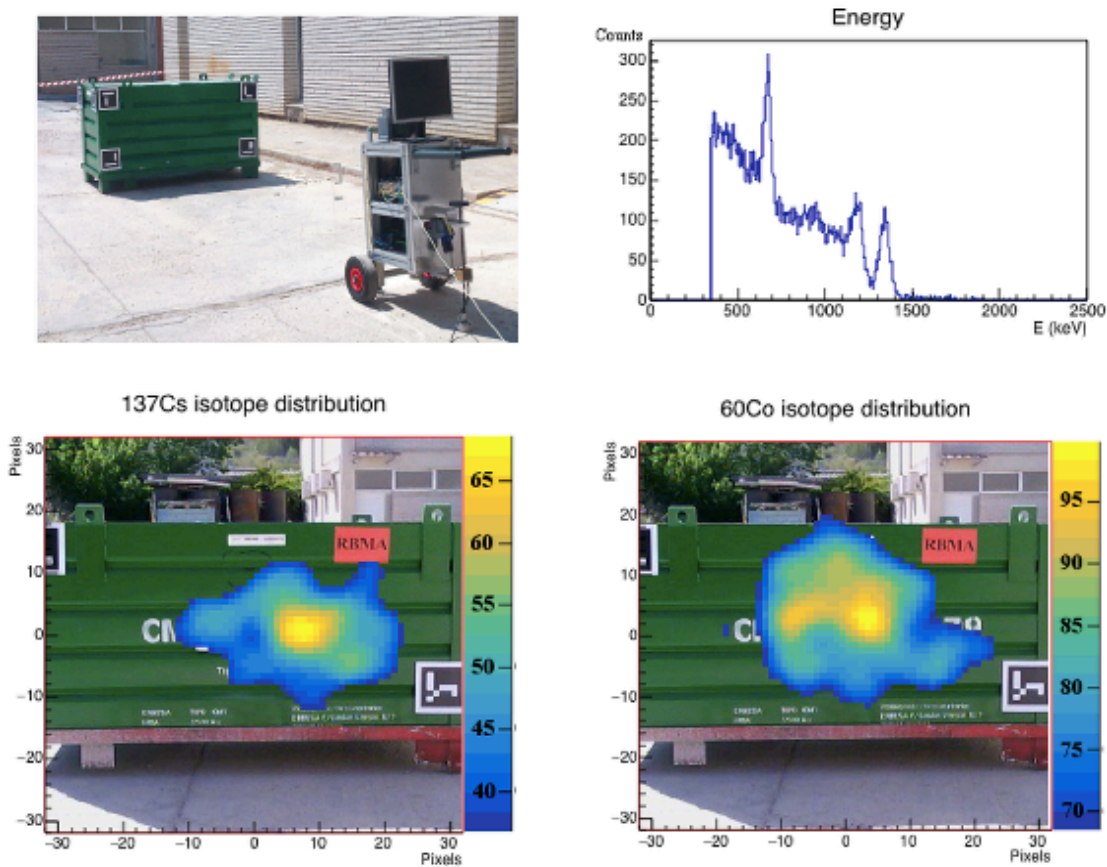

**Fig. 16: GUALI in-field measurement of a container using visual markers to calculate the detector to the (frontal face) container distance (top left). Energy spectrum of the measurement (top right).**

**Gamma images of the container activity distribution obtained with GUALI for $^{137}$Cs (bottom left) and $^{60}$Co (bottom right) radioisotopes (in Bq/g units).**

Another common scenario in NNPs is the inspection of walls or blocks from demolished walls. In such works, the goal is to characterize their radioisotope contamination and decide for their best treatment according to the radionuclides composition and activities found therein. Fig.17 shows the gamma images obtained for the inner face of a concrete piece of the bio-shield used in the reactor vessel. Again, the detector distance to the object was calculated by placing visual markers in the block faces. As it can be observed, the face shown contains contamination of $^{60}$Co, whereas most of the contamination arises from the decay of $^{137}$Cs. The lateral and posterior sides of the block showed no appreciable contamination. This is in agreement with the previously measured surface activity by means of counter detectors. The surface shown in Fig.17 corresponds to the one in contact with the water of the nuclear vessel, whereas the other faces showed much lesser contamination. In this case, the image taken with GUALI allowed to establish a decommissioning approach, whereby the outer layer of the concrete would be removed from the inner side of the block, until successive measurements would reveal a sufficiently low activity in the rest of the block. This will contribute to faster declassify a large mass and volume of material, which otherwise would consume an unnecessary amount of disposal resources and time.

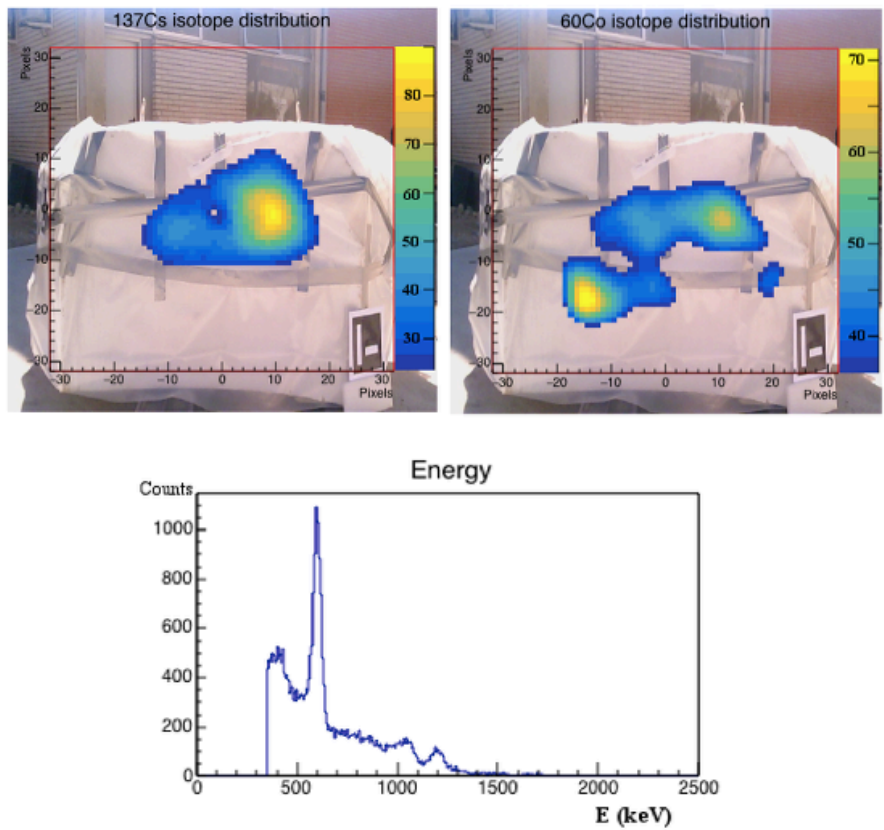

**Fig. 17: Gamma images of the activity distribution in a concrete piece from the reactor vessel for $^{137}$Cs (top left) and $^{60}$Co (top right) radioisotopes (in kBq/pixel units). Energy spectrum is shown in the bottom.**

Finally, we performed a measurement to possibly detect gamma-ray activity escaping from the shielding in the road transportation of drums. In this case, the object to measure was a large truck loaded with one highly active drum surrounded by 8 drums of Low Activity (arranged in a 3x3 array), which were eventually transported to a disposal facility. Since transport is by road from the NPP to the disposal facility, law and regulatory normative demands a very strict shielding level in such a way that there must be public exposure dose levels around the truck. For this reason, drums are placed inside shield container over the truck. As can be shown in Fig.18, we have performed measurements of the lateral side of the truck to check if there was any gamma radiation escaping from the shielding and, in that case, to obtain the corresponding radioisotope identification. None appreciable activity was detected for any radioisotope, which was also in agreement with several measurements made before using HPGe and other inspection detectors. Looking at the radioisotopes gates of $^{137}$Cs and $^{60}$Co (drums radioactive waste composition and activities was known from previous characterizations), we were not able to find any hot spot nor other different activity than the background.

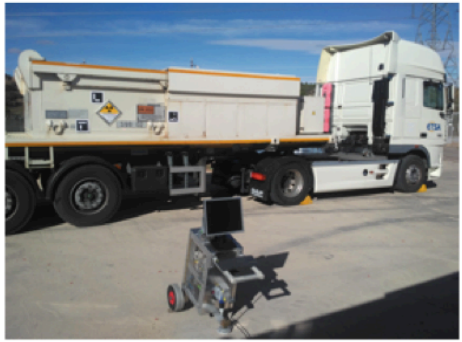

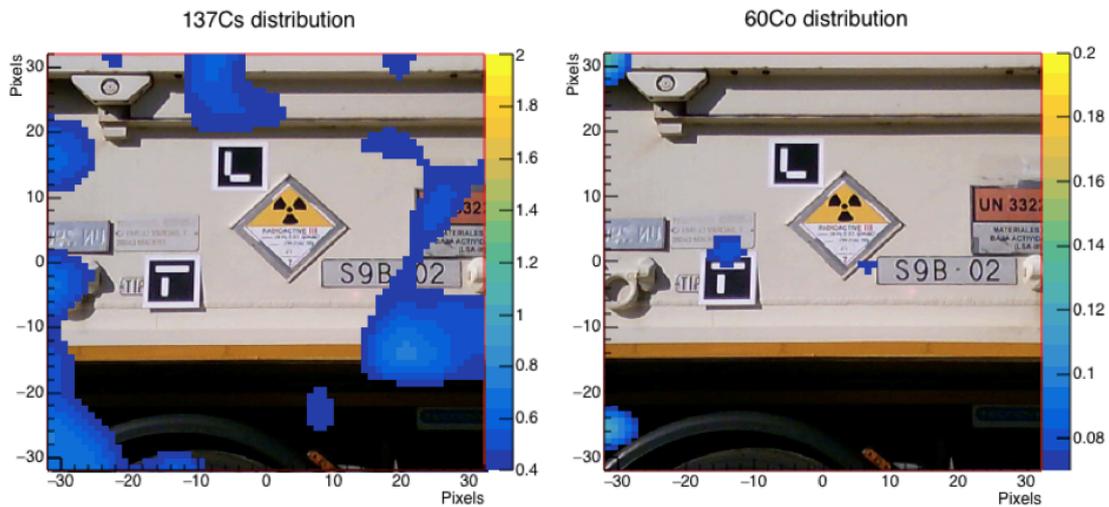

**Fig. 18: Reconstructed gamma images for $^{137}$Cs and $^{60}$Co radioisotopes of a nuclear waste transportation truck (in kBq/pixel units). Energy spectrum is shown in the bottom.**

## 6. Conclusions

In this article, we have given an overview on a new portable system named GUALI (Gamma Unit Advanced Location Imager) intended for high resolution in-field real-time gamma imaging and spectroscopy in Nuclear Power Plants and similar environments.

GUALI represents a complementary diagnostic and monitoring tool, with respect to existing devices. In particular, in comparison with HPGe-based box-counters commonly used for container declassification, the later show an unparalleled spectroscopic and sensitivity performance for the identification of g-ray emitting radioisotopes. On the other hand, employing comparable measuring times and distances, GUALI can contribute to a more accurate waste characterization by means of its imaging capability, relatively high efficiency and real-time operability.

Thanks to its portability and capability to automatically match the visible and gamma images based on the Aruco frames, operating personnel are able to rapidly locate, identify and measure radioactive hot-spots and therefore perform a smarter and more effective characterization according both to the radioisotope spatial distribution. In several situations, the portability and self-management capability of GUALI represents also a key advantage for the effective and versatile inspection of a variety of residues and contaminated objects.

## Acknowledgments

IFIC's mechanics workshop in general, and M. Monserrate in particular for his helpful contribution to the design and realization of the mechanical aspects. Also to Liczandro Hernández and A. González from the I3M for the AMIC2GR support and helpful discussions.